\documentclass[conference]{src/IEEEtran}
\IEEEoverridecommandlockouts
\usepackage{cite}
\usepackage{amsmath,amssymb,amsfonts}
\usepackage{algorithm}
\usepackage{algpseudocode}
\usepackage[nolist]{acronym}
\usepackage{graphicx}
\usepackage{siunitx}
\usepackage{textcomp}
\usepackage{multirow}
\usepackage{arydshln}
\usepackage{caption}
\usepackage{subcaption}
\usepackage{booktabs}
\usepackage{tikz}
\ifCLASSOPTIONcompsoc
\usepackage[caption=false,font=normalsize,labelfont=sf,textfont=sf]{subfig}
\else
\usepackage[caption=false,font=footnotesize]{subfig}
\fi
\usepackage{tikz}

\usepackage{xcolor}

\def\mathlette#1#2{{\mathchoice{\mbox{#1$\displaystyle #2$}}%
		{\mbox{#1$\textstyle #2$}}%
		{\mbox{#1$\scriptstyle #2$}}%
		{\mbox{#1$\scriptscriptstyle #2$}}}}
\def\ve#1{\mathlette{\boldmath}{#1}}

\def\BibTeX{{\rm B\kern-.05em{\sc i\kern-.025em b}\kern-.08em T\kern-.1667em\lower.7ex\hbox{E}\kern-.125emX}}

\makeatletter
\let\old@ps@headings\ps@headings
\let\old@ps@IEEEtitlepagestyle\ps@IEEEtitlepagestyle
\def\confheader#1{%
\def\ps@headings{%
\old@ps@headings%
\def\@oddhead{\strut\hfill#1\hfill\strut}%
\def\@evenhead{\strut\hfill#1\hfill\strut}%
}%
\def\ps@IEEEtitlepagestyle{%
\old@ps@IEEEtitlepagestyle%
\def\@oddhead{\strut\hfill#1\hfill\strut}%
\def\@evenhead{\strut\hfill#1\hfill\strut}%
}%
\ps@headings%
}
\makeatother

\newcommand\submittedtext{%
  \footnotesize This work has been submitted to the IEEE for possible publication. Copyright may be transferred without notice, after which this version may no longer be accessible.}

\newcommand\submittednotice{%
\begin{tikzpicture}[remember picture,overlay]
\node[anchor=south,yshift=10pt] at (current page.south) {\fbox{\parbox{\dimexpr0.65\textwidth-\fboxsep-\fboxrule\relax}{\submittedtext}}};
\end{tikzpicture}%
}

\newcommand\copyrighttext{%
  \footnotesize \textcopyright \the\year{} IEEE. Personal use of this material is permitted. Permission from IEEE must be obtained for all other uses, including reprinting/republishing this material for advertising or promotional purposes, collecting new collected works for resale or redistribution to servers or lists, or reuse of any copyrighted component of this work in other works.}

\title{Passive Channel Charting: Locating Passive Targets using a UWB Mesh}

\def\authorrefmark#1{\ensuremath{^{\textbf{#1}}}}
\author{
    Raffael Poeggel\authorrefmark{1},
    Maximilian Stahlke\authorrefmark{1},
    Jonas Pirkl\authorrefmark{1},
    Jonathan Ott\authorrefmark{1},
    George Yammine\authorrefmark{1},\\
    Tobias Feigl\authorrefmark{1,}\authorrefmark{2},
    and Christopher Mutschler\authorrefmark{1}\\
    {\tt\footnotesize\{raffael.poeggel, maximilian.stahlke, jonas.pirkl, jonathan.ott, george.yammine,}\\ 
    {\tt\footnotesize tobias.feigl, christopher.mutschler\}@iis.fraunhofer.de} 
    \vspace{+1mm}\\ 
    \IEEEauthorblockA{\authorrefmark{1}
        Fraunhofer Institute for Integrated Circuits (IIS), 
        Division Positioning and Networks, 
        90411 Nürnberg, Germany
        }
    \IEEEauthorblockA{\authorrefmark{2}
        Friedrich-Alexander-Universität Erlangen-Nürnberg (FAU), 
        Programming Systems Group,
        91052 Erlangen, Germany 
    }
}

\begin{document}
\maketitle
\submittednotice

\begin{acronym}[tdma]
    \acro{csi}[CSI]{channel state information}
    \acro{uwb}[UWB]{ultra-wideband}
    \acro{rssi}[RSSI]{radio strength indicator}
    \acro{los}[LOS]{line-of-sight}
    \acro{nlos}[NLOS]{non-line-of-sight}
    \acro{pdr}[PDR]{pedestrian dead reckoning}
    \acro{cc}[CC]{channel charting}
    \acro{fp}[FP]{fingerprinting}
    \acro{mpc}[MPC]{multi path component}
    \acro{rti}[RTI]{radio tomography imaging}
    \acro{dfl}[DFL]{device free localization}
    \acro{pcc}[PCC]{passive channel charting}
    \acro{imu}[IMU]{interial measurement unit}
    \acro{tdma}[TDMA]{time division multiple access}
    \acro{pca}[PCA]{principal component analysis}
    \acro{mae}[MAE]{Mean average error}
    \acro{cir}[CIR]{Channel impulse response}
    \acro{mse}[MSE]{mean squared error}
\end{acronym}

\begin{abstract}
Fingerprint-based passive localization enables high localization accuracy using low-cost UWB IoT radio sensors. However, fingerprinting demands extensive effort for data acquisition. The concept of channel charting reduces this effort by modeling and projecting the manifold of \ac{csi} onto a 2D coordinate space. So far, researchers only applied this concept to active radio localization, where a mobile device intentionally and actively emits a specific signal.

In this paper, we apply channel charting to passive localization. We use a pedestrian dead reckoning (PDR) system to estimate a target's velocity and derive a distance matrix from it. We then use this matrix to learn a distance-preserving embedding in 2D space, which serves as a fingerprinting model. In our experiments, we deploy six nodes in a fully connected ultra-wideband (UWB) mesh network to show that our method achieves high localization accuracy, with an average error of just 0.24\,m, even when we train and test on different targets.
\end{abstract}


\section{Introduction}

The widespread deployment and rapid expansion of IoT radio devices and infrastructure, such as WiFi and \ac{uwb}, have transformed indoor environments into dense, radio-rich spaces~\cite{zhu_indoor_2020}. At the same time, device-free, passive radio sensing enables the tracking of objects or individuals without requiring them to carry any devices~\cite{youssef_challenges_2007}. Together, these developments create strong potential for a wide range of passive sensing applications, including monitoring, surveillance, presence detection, activity recognition, and even sensitive tasks like vital sign monitoring~\cite{ chen_wi-fi_2023, cheraghinia_comprehensive_2024}. Such applications are especially relevant in areas like security~\cite{lai_through-wall_2005}, personal health~\cite{gao_vital_2005}, and elderly care~\cite{li_uwb-radar-based_2024}. 

Most related work relies on \ac{rssi} measurements from various radio systems such as WiFi~\cite{ma_wifi_2020}, Bluetooth~\cite{zhang_comprehensive_2013}, and \ac{uwb}\cite{cheraghinia_comprehensive_2024}. However, \ac{rssi} provides limited environmental information, resulting in restricted sensing accuracy\cite{yang_rssi_2013}. To overcome this, recent approaches use channel state information (\ac{csi}), which offers finer-grained insights into environmental effects such as reflections, scattering, and shadowing~\cite{liu_survey_2019, suroso_wi-fi_2023, yang_rssi_2013, ma_wifi_2020}. This richer data enables higher accuracy, especially with high-bandwidth systems such as \ac{uwb}~\cite{cimdins2022exploiting}. For passive localization, analytic, i.e., non-data-driven methods extract reflections from a target, e.g., a person or automated guided vehicle~\cite{chang2010people, bocus_passive_2021}, or model the target’s impact on multipath propagation~\cite{gentner2023ranging}. However, these methods typically require \ac{los} or detailed environmental models, limiting their applicability in cluttered or complex spaces. To enable localization in \ac{nlos} environments, \ac{fp} can be used~\cite{shi_accurate_2018}. Unlike analytic methods, \ac{fp} maps recorded \ac{csi} directly to positions, making it robust to \ac{nlos} conditions and environmental complexity. However, \ac{fp} requires costly data acquisition with reference positions~\cite{rao_novel_2024}. Environmental changes, e.g., movement of large objects, invalidate fingerprints, require updated reference data, and raise lifecycle management costs~\cite{stahlke2022transfer}. 
Recently, the concept of \ac{cc} was proposed to reduce or eliminate the need for reference data~\cite{studer2018channel}, enabling simpler deployment and easier updates of \ac{fp} models. \ac{cc} models the manifold of \ac{csi} and maps it to 2D coordinates using only radio signals, either in a self-supervised manner~\cite{stahlke2023indoor} or with additional context, such as physical motion assumptions~\cite{euchner2022improving} or low-cost sensor data~\cite{stahlke2024velocity}. However, prior work only applied \ac{cc} to active localization with mobile transmitters.

Alongside Euchner et al.~\cite{euchner2025passive}, we propose the concept of \ac{pcc} for passive localization. While they apply \ac{pcc} using a distributed Wi-Fi multi-antenna setup, we focus on a fully connected UWB mesh. Our approach localizes a mobile target based on its influence on environmental reflections and direct signal reflections. Due to the smooth movement of the target, the resulting variations in the radio signals remain smooth, allowing us to model the \ac{csi} manifold and project it into spatial coordinates, analogous to active \ac{cc}.
We implement velocity-based \ac{cc}~\cite{stahlke2024velocity} that estimates the target's velocity using a \ac{pdr} system. Our experiments demonstrate that we achieve high localization accuracy, with a mean absolute error (MAE) of up to 0.24\,m, even when training and testing on different individual targets. We also study the mesh configuration to provide insights for optimal system design.

The rest of this paper is structured as follows.
Sec.~\ref{sec:related_work} discusses related work.
Then, Sec.~\ref{sec:method} introduces our method.
Sec.~\ref{sec:experimental_setup} describes our experimental setup.
Sec.~\ref{sec:evaluation} discusses the results, before Sec.~\ref{sec:conclusion} concludes.

\begin{figure*}[!tp]
    \centering
    \includegraphics[width=\textwidth]{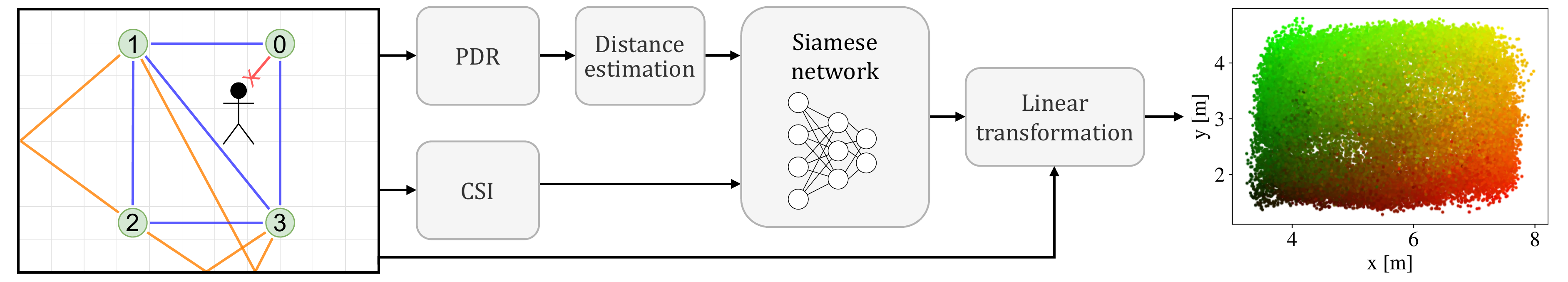}
    \vspace{-0.2cm}
    \caption{Our Passive Channel Charting (PCC) pipeline records \ac{csi} from static \ac{uwb} nodes and distances from an IMU attached to a moving target. It computes a sparse distance matrix between \ac{csi} measurements. A Siamese network learns a distance-preserving embedding from the high-dimensional \ac{csi} space to 2D spatial coordinates. To align the \ac{cc} (isometric) coordinates with the real-world radio geometry, \ac{pcc} applies a linear transformation based on few labeled data-points.}
    \label{fig:pos_pipeline}
    \vspace{-0.45cm}
\end{figure*}

\section{Related Work}%
\label{sec:related_work}%

\textbf{Active localization algorithms} such as classical Round-Trip Time (RTT) and Time Difference of Arrival (TDoA) multilateration~\cite{cao_indoor_2020,kim_efficient_2015} perform worse than \ac{fp} methods in multipath-rich environments~\cite{feng_wifi_2023}. However, \ac{fp} methods need \ac{csi} databases with reference positions, which increase costs~\cite{gidey_ohettlal_2022}. Unsupervised approaches solve this problem and remove the need for labeled training data~\cite{hutchison_dccla_2013}. Studer et al.\cite{studer2018channel} introduce \ac{cc} as an unsupervised alternative to traditional \ac{fp}. \ac{cc} models the \ac{csi} manifold and maps it to 2D coordinates through manifold-preserving dimensionality reduction. This enables fully unsupervised \ac{fp} and captures environmental reflections. Many \ac{cc} methods use distance matrices to represent spatial relationships in \ac{csi} data\cite{stephan2024angle, euchner2024leveraging}. Although \ac{cc} lacks global consistency, Stahlke et al.\cite{stahlke2023indoor} achieve it with geodesic distances derived from \ac{csi}. Self-supervised \ac{cc} reaches moderate localization accuracy\cite{studer2018channel,le2021efficient}, but additional sensor data improves results. Low-cost sensors such as IMU or odometry estimate relative velocities. They provide inter-sample distances and help \ac{cc} match the accuracy of supervised fingerprinting~\cite{stahlke2024velocity}.

\textbf{Passive localization algorithms} locate people by using signal reflections and geometric models. Sensor noise reduces positioning accuracy. Zhang et al.\cite{zhang_rf-based_2007} study geometric methods that reduce the effect of noise from a WiFi RSSI sensor array in \ac{dfl}. Their methods support both single- and multi-target tracking. Ledergerber et al.\cite{ledergerber_multi-static_2020} use high-bandwidth UWB systems to detect reflections and measure delays in CIR for outdoor tracking. Bocus et al.\cite{bocus_passive_2021} adapt this to indoor environments. They remove clutter to isolate reflections from a target. Gentner et al.\cite{gentner2023ranging} use environmental reflections, assign \acp{mpc} to virtual anchors, and solve localization with geometric models. These  methods often require \ac{los} or detailed maps, which limits their use in cluttered areas. Zhang et al.\cite{zhang_variance-constrained_2024} improve accuracy in complex environments. They divide space based on statistical features, train local \ac{dfl} models, and merge them with a variance-constrained method. Rao et al.\cite{rao_novel_2024} propose a low-complexity method. They use calibrated \ac{csi} phases from a single link to run \ac{dfl} and reduce hardware costs. Their approach depends on earlier phase calibration~\cite{rao_dfphasefl_2020} and manual tuning. Recent unsupervised \ac{dfl} methods use probabilistic RSSI models to estimate presence and position. Al-Husseiny et al.\cite{al-husseiny_unsupervised_2019} apply a hidden Markov model. Hillyard et al.\cite{hillyard_never_2020} extend this with a Bayesian method that adapts to changes. Kaltiokallio et al.~\cite{kaltiokallio_unsupervised_2021} use Gaussian smoothing to increase accuracy and efficiency. These methods rely on iterative inference and do not scale well. They also depend on \ac{los} \ac{rssi} links, which require dense sensor deployments and remain vulnerable to \ac{mpc} interference.

Unlike all existing approaches, we draw inspiration from \ac{cc} to achieve accurate localization of moving people by combining radio \ac{csi} with additional information such as odometry measurements, and we integrate these ideas with the concept of passive localization. 

\section{Methodology}%
\label{sec:method}%

\ac{csi} dependents on the surrounding environment due to effects such as reflections, scattering, and blockage, and it changes smoothly w.r.t. the movement of the transmitter. As a result, the radio signals lie on a low-dimensional manifold corresponding to positions in space. \ac{cc} leverages this property to create a mapping from high-dimensional \ac{csi} data to spatial coordinates~\cite{studer2018channel}. However, existing \ac{cc} approaches have only been applied in active localization scenarios, where a target carries a transmitting device.

In passive localization, a target does not carry any active transmitter. Instead, the \ac{csi} varies due to the target's influence on the radio propagation, either by blocking existing paths or acting as a reflector, thereby affecting the signals in the environment. Due to the target's smooth movement, the \ac{csi} recorded by the static infrastructure also changes smoothly. Hence, as in active \ac{cc}, the \ac{csi} in passive setups lies on a spatial manifold. We extend this concept by projecting the manifold of static \ac{csi} measurements to spatial positions, an approach we refer to as passive channel charting (PCC).

Fig.~\ref{fig:pos_pipeline} illustrates the end-to-end positioning pipeline of our PCC method. We deploy a static UWB mesh in the environment, while a target moves within it, influencing the radio signals by blocking and reflecting them. Most \ac{cc} algorithms estimate pairwise distances between measurements and learn a 2D embedding that preserves those distances. Following the approach of Stahlke et al.\cite{stahlke2024velocity}, we estimate a distance matrix based on the target’s velocity, which we obtain using a \ac{pdr} system attached via an \ac{imu}~\cite{hou2020pedestrian}. We compute the distance between measurements by integrating the velocity over time:
\begin{equation}
    \label{eq:vel_dist}
    d_{n,k} = \left\| \int_{t=t_n}^{t_k} v(t)\, \mathrm{d}t  \right\|_2^{} \;,
\end{equation}
where $v(t)$ is the estimated velocity of the target at time $t$ and $d_{n,k}$ the distance between measurements $n$ and $k$. To reduce the impact of \ac{pdr} drift, we limit the integration window, resulting in a sparse distance matrix. 

To learn the underlying manifold, we use a Siamese neural network~\cite{lei2019siamese} that embeds high-dimensional \ac{csi} into 2D coordinates while preserving the estimated distances. The loss function for training is defined as:
\begin{equation}
\label{eq:dist_loss}
    \mathcal{L}_\mathsf{d} = \beta (d_{n,k}-\|f_{\theta}(\ve{x}_n)^{}-f_{\theta}(\ve{x}_k)^{}\|_2^{} ) ^2,
\end{equation}
where $f_{\theta}(\cdot)$ is the neural network parameterized by $\theta$, $\ve{x}$ represents the input \ac{cir} vectors, $||\cdot||_2$ is the Euclidean distance, and $\beta$ is a weighting factor. Shorter distances are weighted more heavily to account for increasing \ac{pdr} drift over longer time spans. As the learned embedding preserves only relative distances, we transform the resulting local \ac{cc} coordinate frame to the global coordinate frame using a small set of \ac{csi} samples with known positions. Similar to~\cite{stahlke2023indoor}, we estimate \ac{cc} coordinates for these reference samples and compute a linear transformation to the global positions via least squares estimation.

\section{Experimental Setup}%
\label{sec:experimental_setup}%

This Section describes the \ac{uwb} mesh (Sec.~\ref{ssec:experimental_setup:mesh}), the \ac{pdr} system (Sec.~\ref{ssec:experimental_setup:pdr}), the Siamese Network configuration (Sec.~\ref{ssec:experimental_setup:siamese}), and the datasets (Sec.~\ref{ssec:experimental_setup:datasets}).

\subsection{\ac{uwb} Mesh}%
\label{ssec:experimental_setup:mesh}%

The system operates at a center frequency of 4.5\,GHz (bandwidth = 499.99\,MHz) with 6 DW1000-based nodes forming a mesh network. 4 nodes (0–3) act as transceivers, and 2 nodes (4, 5) act as receivers, forming a multi-static radar setup with 14 total links. A \ac{tdma} protocol assigns non-overlapping time slots for transmissions. The system records the \ac{cir} for each link and limits it to 90\,\si{ns} (90 samples at~1\,\si{ns} resolution), covering about 30\,m of multipath propagation. The system records CIRs at 40\,\si{Hz}.

\subsection{\ac{pdr}}%
\label{ssec:experimental_setup:pdr}%

An IMU (XSense MTI 300) attached to the foot estimates a person's velocity. We apply the Gaitmap toolset~\cite{kuderle2024gaitmap} for velocity estimation. As the IMU is mounted in a random orientation, we align roll and pitch using gravity and correct yaw with a principal component analysis. For trajectory estimation, we use angular velocity, acceleration, and magnetic field. We apply an Error-State Kalman Filter~\cite{sola2012quaternion} together with Rauch-Tung-Striebel smoothing~\cite{colomar2012smoothing} to estimate the target's velocity.

\subsection{Siamese network architecture}%
\label{ssec:experimental_setup:siamese}%

We use a ResNet-based Siamese network with 5.6\,M parameters. Tab.~\ref{tab:nn-config} lists the full configuration, that includes all UWB links, i.e., number of antennas. The input dimensions vary depending on the setup. We start with a 1$\times$1 CNN layer to expand the channel dimension from 1 to 64. We then apply 9 ResNet blocks. After each ResNet block group (3 blocks), the network doubles the channel dimension and reduces the time and antenna dimensions with a stride of 2. Each CNN layer uses a kernel size of 3 and padding of 1. Global average pooling follows across the antenna and time dimensions, leaving a 256-dimensional channel vector. 2 linear layers with ReLU reduce this to the 2D target position.

\begin{table}[t]
    \centering
    \caption{Network configuration (output shapes / layer).}
    \begin{tabular}{lr}
        Layer Type       & (time $\times$ antenna $\times$ channels)\\ \hline
        Input            & 90 $\times$ 14 $\times$ \quad 1                               \\
        CNN              & 90 $\times$ 14 $\times$ \, 64                              \\
        ResNet block group      & 90 $\times$ 14 $\times$ \, 64                              \\
        ResNet block group      & 44 $\times$ \, 6 $\times$ 128                              \\
        ResNet block group      & 21 $\times$ \, 2 $\times$ 256                              \\
        GlobalAvgPooling &  256                               \\
        LinearLayer      &  200                               \\
        LinearLayer      &  2                                 \\
    \end{tabular}
    \label{tab:nn-config}
    \vspace{-0.45cm}
\end{table}

\subsection{Datasets} 
\label{ssec:experimental_setup:datasets}

\begin{figure}[!bp]
    \vspace{-0.45cm}
    \centering
    \begin{minipage}[b]{.49\columnwidth}
        \centering
        \includegraphics[width=\columnwidth]{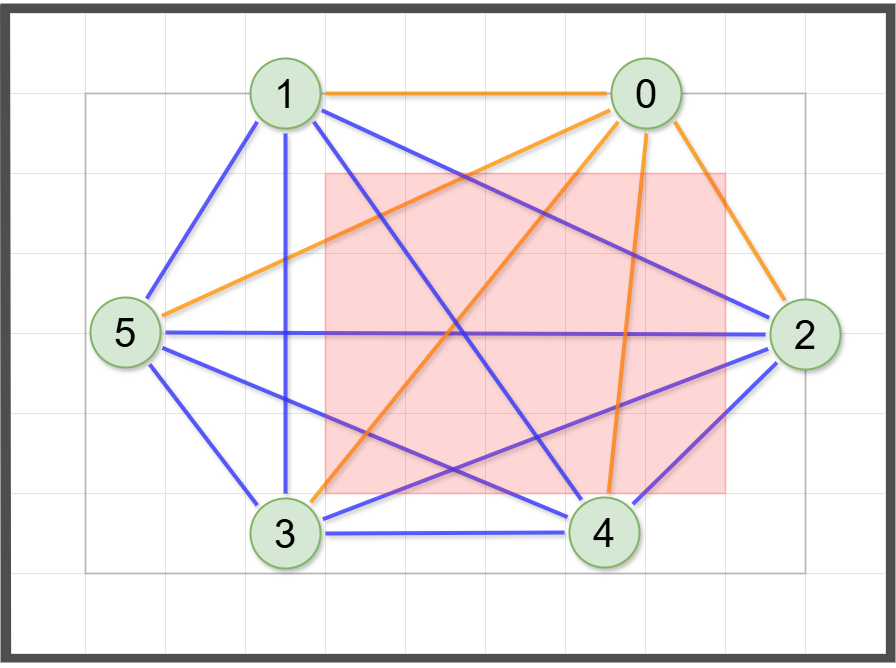}
    \end{minipage}\hfill
    \begin{minipage}[b]{.49\columnwidth}
        \centering
        \includegraphics[width=\columnwidth]{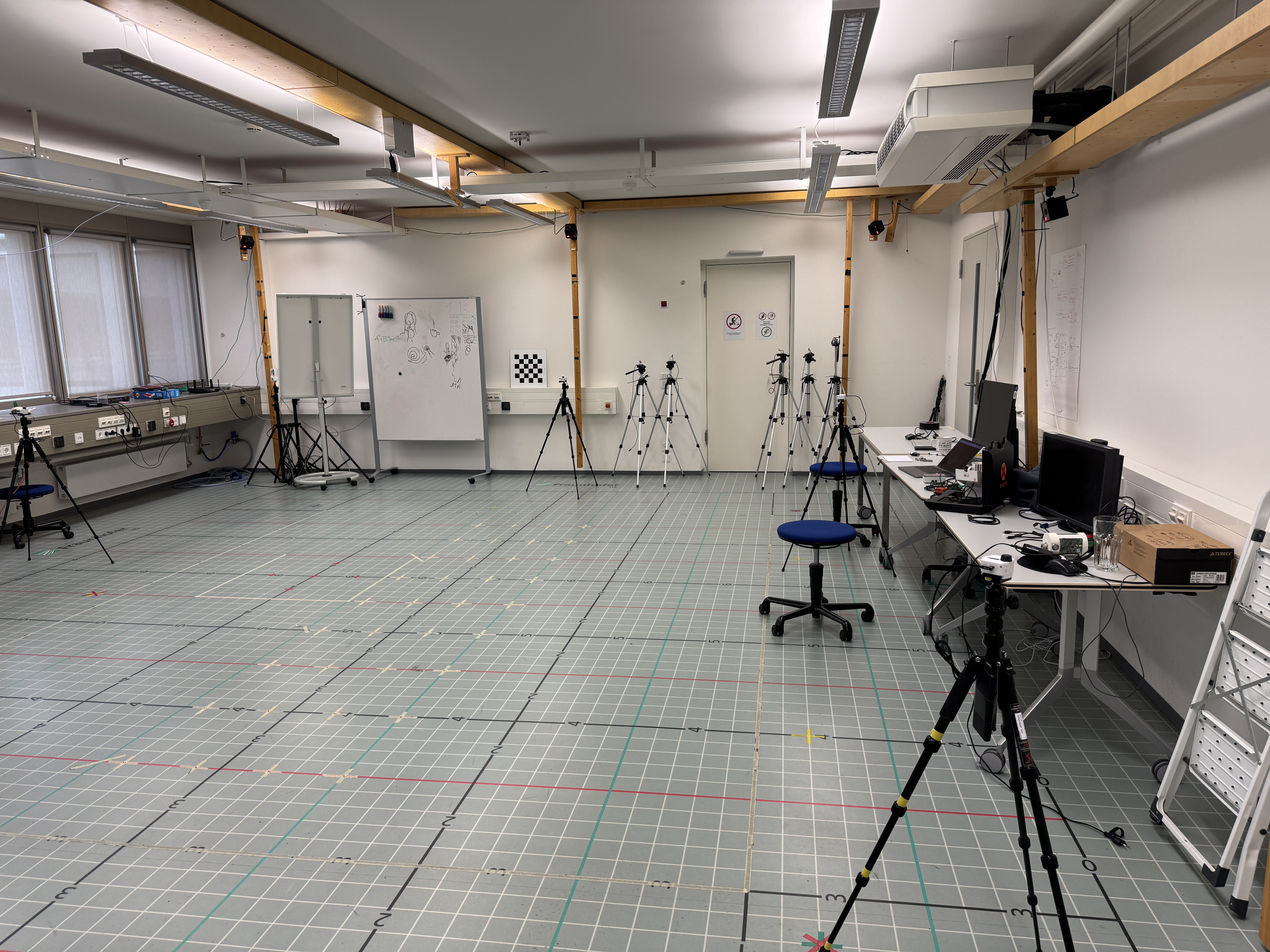}
    \end{minipage}
    \vspace{-0.2cm}
    \caption{
    Top-down schematic (left) and photo (right) of the environment (5m $\times$ 4m, red area), with: reflective surfaces, e.g., windowsills, whiteboards, monitors, and windows; UWB nodes (green) at 1.5m height; LoS links in the multi-static setup (blue lines); bi-static setup (orange lines).}
    \label{fig:environment}
\end{figure}

We evaluate our algorithm in an office-like environment measuring 11\,m $\times$ 8\,m, see Fig.~\ref{fig:environment}. The room contains office equipment such as whiteboards, tables, and reflective walls creating multipath. The fully connected mesh has 14 direct links, shown as solid lines; orange lines highlight links to node 0. The red area marks the walking zone. To study generalization, we record data from 3 people. For training, the first target wears the PDR system, see Sec.~\ref{ssec:experimental_setup:pdr}. Others do not wear PDR and serve for testing only. All walk at a slow pace (1\,m/s) on random paths. The training set includes 40\,min of CSI and velocity data, totaling about 70,000 samples. The test sets include 20\,min (P1; 35,000 samples) and 10\,min (P2; 17,500 samples). An optical reference system provides reference positions at sub-centimeter accuracy. To use the power delay profiles of the \acp{cir} in a neural network, we stack them into a 2D array with shape $[N, 90]$, where $N$ is the number of \acp{cir} and $90$ is the number of samples per CIR. We normalize the input to $[0,1]$ using the maximum magnitude from the training set.

\begin{figure}[!tp]
  \includegraphics[width=1.0\columnwidth]{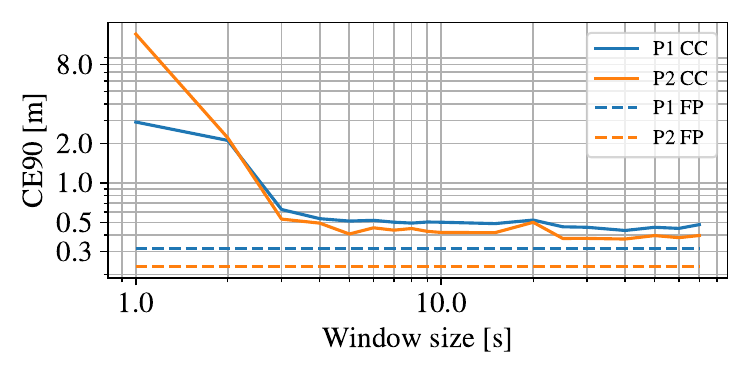}
  \caption{Localization error (CE90) of \ac{pcc} for different window sizes on datasets from P1 (blue, solid) and P2 (orange, solid). Dashed lines show results for supervised FP.}
  \label{fig:res_window_size}
\end{figure}

\section{Evaluation}%
\label{sec:evaluation}%

We evaluate our \ac{pdr}-based \ac{pcc} for different \ac{pdr} window sizes (Sec.~\ref{sec:eval_PDR_CC}) and analyze how the mesh composition affects localization accuracy (Sec.~\ref{sec:eval_ablation}). 

As an upper bound, we compare our method to supervised \ac{fp}\cite{shi_accurate_2018}. The \ac{fp} model uses the same Siamese network architecture (Sec.~\ref{ssec:experimental_setup:siamese}) and directly fits the output to the reference positions of the \ac{csi} values using \ac{mse} loss. We measure accuracy using \ac{mae} and the 90th percentile of the error distribution (CE90), where the error is the Euclidean distance between the predicted and reference positions. Each model trains for 500 epochs using the AdamW optimizer~\cite{loshchilov2019decoupledweightdecayregularization} with a learning rate of 0.0001 and a batch size of 128. A learning rate scheduler reduces the learning rate by factor 2 if the training does not improve for 50 consecutive epochs. We train the models on data from 1 target and test on 2 others (P1, P2) to evaluate generalization. We fit a linear transformation to map the 2D manifold of the \ac{pcc} method into absolute coordinates, for which we use 20 reference positions from the training set.

\subsection{\ac{pdr}-based \ac{pcc}}%
\label{sec:eval_PDR_CC}%

We evaluate the \ac{pdr}-based \ac{pcc} for different window sizes on the fully connected UWB mesh, where all node links contribute to \ac{pcc}. Tab.~\ref{tab:res_CC_window} shows the results; Fig.~\ref{fig:res_window_size} visualizes them for P1 (blue) and P2 (orange) with solid lines. As an lower bound, we include supervised FP results as dashed lines.

Supervised FP achieves highest accuracy: CE90 = 0.31\,m and MAE = 0.17\,m for P1; CE90 = 0.23\,m and MAE = 0.13\,m for P2. The small 0.08\,m gap between P1 and P2 shows strong generalization. \ac{pcc} shows a similar pattern, with an average error difference of 0.07\,m across window sizes, enabling target-independent localization.

\ac{pcc} reaches its best accuracy at a 40\,s window: CE90 = 0.43\,m and MAE = 0.21\,m for P1; CE90 = 0.37\,m and MAE = 0.26\,m for P2. This results in a CE90 difference of 0.12\,m for P1 and 0.14\,m for P2 compared to supervised FP. Window size strongly impacts accuracy. Small windows (1–2\,s) produce poor results: CE90 = 2.11\,m, MAE = 1.43\,m for P1; CE90 = 2.21\,m, \ac{mae} = 1.29\,m for P2.

\begin{figure*}[!tp]
    \centering
    \begin{minipage}[b]{.49\columnwidth}
        \centering
        \includegraphics[width=\columnwidth]{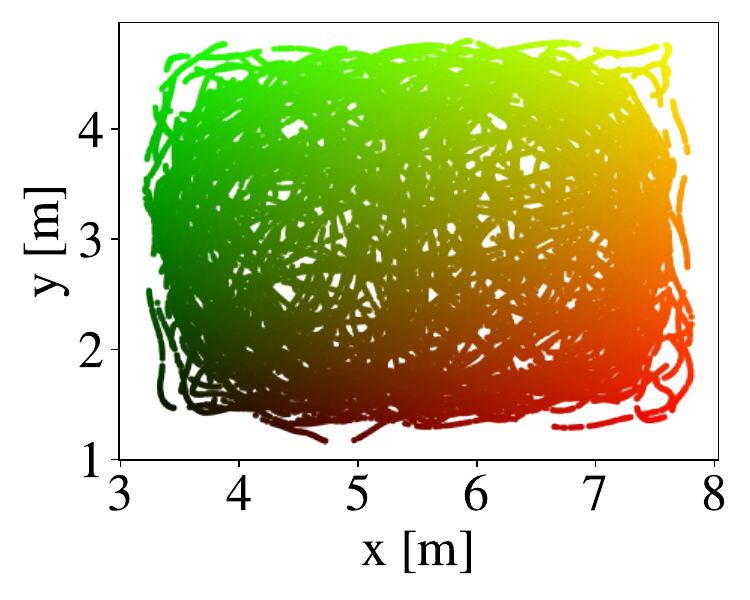}
    \end{minipage}\hfill
    \begin{minipage}[b]{.49\columnwidth}
        \hspace{-20pt} 
        \centering
        \includegraphics[width=\columnwidth]{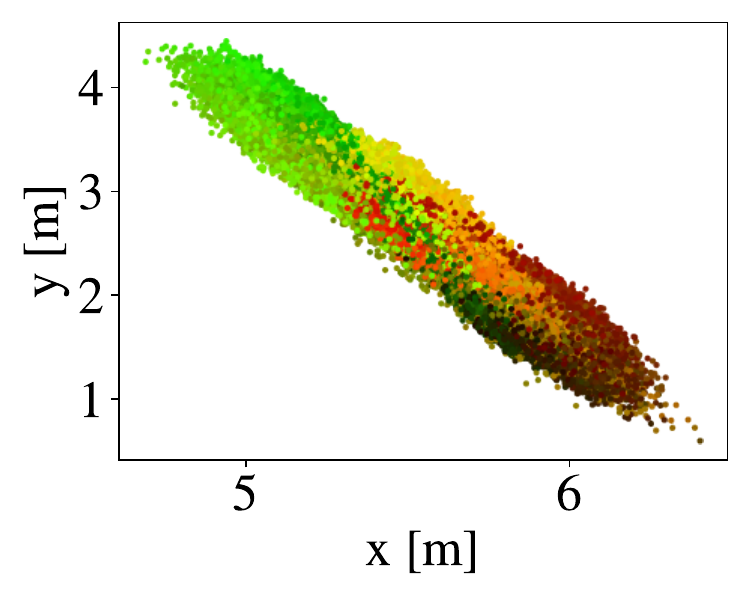}
    \end{minipage}\hfill
    \begin{minipage}[b]{.49\columnwidth}
        \hspace{-28pt} 
        \centering
        \includegraphics[width=\columnwidth]{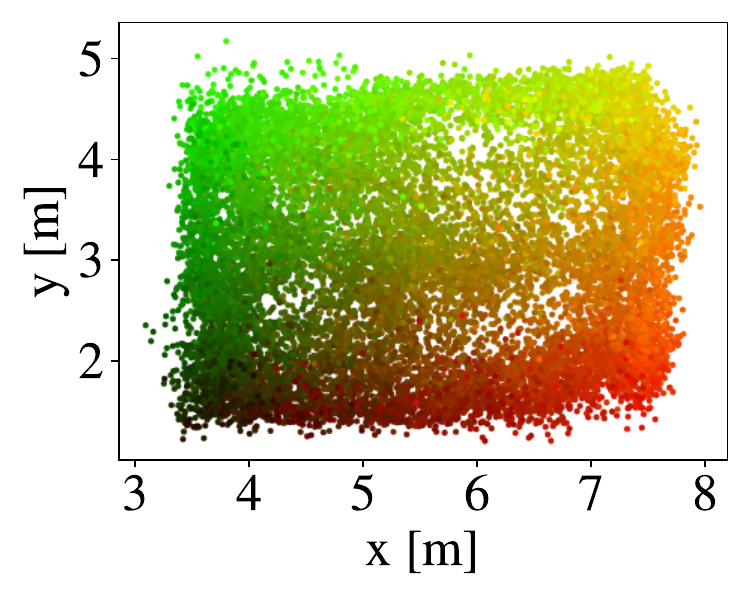}
    \end{minipage}
    \begin{minipage}[b]{.49\columnwidth}
        \hspace{-20pt} 
        \centering
        \includegraphics[width=\columnwidth]{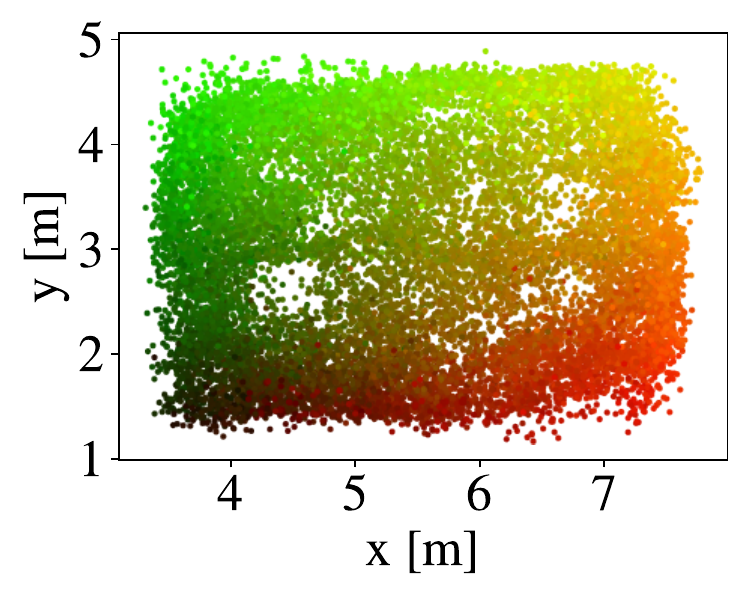}
    \end{minipage}
    \caption{Results. Left: reference trajectory of P1 with a color gradient showing spatial neighborhoods; second: channel chart with a 2\,s window; third: 10\,s window; right: 40\,s window.}
    \label{fig:cc_window_sizes}
\end{figure*}

Fig.~\ref{fig:cc_window_sizes} shows the channel charts for P1 (left: reference trajectory). A color gradient indicates spatial neighborhood. The 2\,s chart is second, the 10\,s chart is third, and the 40\,s chart is rightmost. The 2\,s chart fails to recover the global geometry, though it preserves local structure. This matches prior findings~\cite{stahlke2024velocity}, where velocity-based integration needs large trajectories to capture the full environment. At ~1\,m/s walking speed, 2\,s trajectories span only ~2\,m, which is too short for the 5\,m$\times$4\,m space. Larger windows, such as 10\,s and 40\,s, create globally consistent charts, with no major gain beyond 10\,s.

\begin{table}[!tp]
    \centering
    \setlength\tabcolsep{1.75pt}
    \caption{CE90 localization error of \ac{pcc} for window sizes on data from targets P1 and P2 using the full \ac{uwb} mesh.}
    \begin{tabular}{ll|c|c|c|c|c|c|c|c|c|c|c|c}
    \toprule
    & & \multicolumn{12}{c}{Window size [s]} \\
    \cmidrule(lr){3-14}
     &  & 1& 2& 3& 4& 5& 10& 20& 30& 40& 50& 60& 70\\
    \midrule
    \multirow{2}{*}{P1} & CE90& 2.92& 2.11& 0.63& 0.53& 0.51& 0.50& 0.52& 0.46& \textbf{0.43}& 0.46& 0.45& 0.48\\
 & \ac{mae} & 1.76& 1.43& 0.35& 0.30& 0.29& 0.28& 0.29& 0.26& \textbf{0.25}& 0.26& 0.25& 0.27\\
         \midrule
    \multirow{2}{*}{P2} &  CE90& 13.66& 2.21& 0.53& 0.49& 0.41& 0.42& 0.50& 0.38& \textbf{0.37}& 0.40& 0.38& 0.40\\
 & \ac{mae} & 6.90& 1.29& 0.30& 0.28& 0.23& 0.24& 0.28& 0.21& \textbf{0.21}& 0.22& 0.22& 0.22\\
    \bottomrule
    \end{tabular}
    \label{tab:res_CC_window}
\end{table}

\begin{table*}[!tp]
    \centering
    \setlength\tabcolsep{3pt}
    \caption{Localization errors (CE90 and \ac{mae}) for \ac{uwb} configurations (numbers indicate nodes): ``full'' mesh, $st\_[0,3]$ are single-transmitter setups, $ac\_5$ = [0,1,3,4,5], $ac\_4$ = [4,3,1,0], $ac\_3$ = [3,1,0], $ac\_2\_s$ = [3,0], $ac\_2\_l$ = [2,5].}
    \begin{tabular}{ll|cc|cc|cc|cc|cc|cc|cc|cc|cc|cc}
        \toprule
        & Scenario & \multicolumn{2}{c}{full} & \multicolumn{2}{c}{st\_0}& \multicolumn{2}{c}{st\_1}& \multicolumn{2}{c}{st\_2}& \multicolumn{2}{c}{st\_3}& \multicolumn{2}{c}{ac\_5}& \multicolumn{2}{c}{ac\_4}& \multicolumn{2}{c}{ac\_3}& \multicolumn{2}{c}{ac\_2\_s}& \multicolumn{2}{c}{ac\_2\_l}\\
        \cmidrule(lr){2-22}
        & Target & P1 & P2 & P1 & P2 & P1 & P2 & P1 & P2 & P1 & P2 & P1 & P2 & P1 & P2 & P1 & P2 & P1 & P2 & P1 & P2\\
        \midrule
        \multirow{2}{*}{CC} & \ac{mae} & 0.28 & 0.24 & 0.29 & 0.24 & 0.46 & 0.40 & 0.32 & 0.30 & 0.30 & 0.29 & 0.31 & 0.27 & 0.32 & 0.28 & 0.39 & 0.34 & 0.43 & 0.38 & 0.89 & 0.75\\
        & CE90 & 0.50 & 0.42 & 0.55 & 0.44 & 0.83 & 0.73 & 0.61 & 0.57 & 0.56 & 0.52 & 0.55 & 0.49 & 0.59 & 0.52 & 0.75 & 0.66 & 0.85 & 0.74 & 1.70 & 1.47\\
        \midrule
        \multirow{2}{*}{\ac{fp}} & \ac{mae} & 0.17 & 0.13 & 0.22 & 0.16 & 0.34 & 0.28 & 0.24 & 0.21 & 0.22 & 0.19 & 0.22 & 0.17 & 0.25 & 0.20 & 0.34 & 0.26 & 0.35 & 0.29 & 0.99 & 0.87\\
        & CE90 & 0.31 & 0.23 & 0.40 & 0.28 & 0.71 & 0.61 & 0.47 & 0.42 & 0.39 & 0.37 & 0.41 & 0.31 & 0.49 & 0.39 & 0.72 & 0.52 & 0.69 & 0.57 & 2.40 & 2.10\\
        \bottomrule
    \end{tabular}
    \label{tab:results_ablation}
\end{table*}

\subsection{Ablation study} \label{sec:eval_ablation}

\begin{figure*}[t]
    \centering
    \begin{minipage}[b]{.49\columnwidth}
        \centering
        \includegraphics[width=\columnwidth]{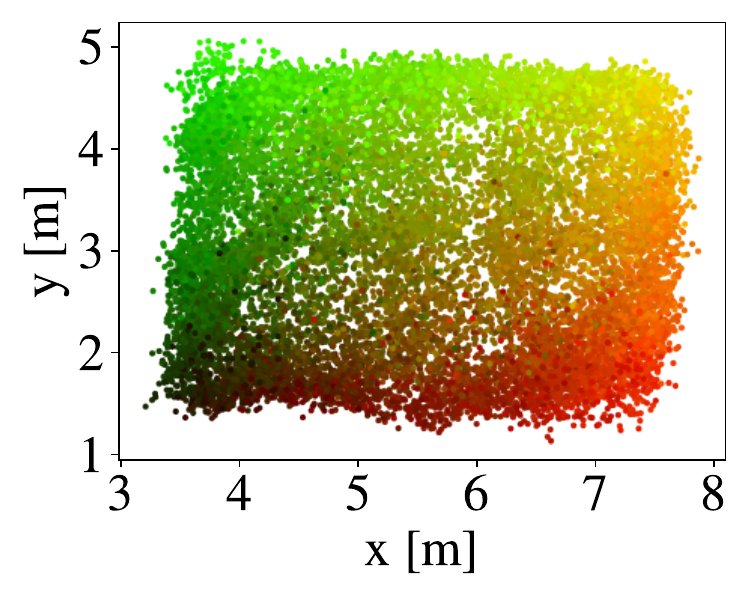}
    \end{minipage}\hfill
    \begin{minipage}[b]{.49\columnwidth}
        \hspace{-20pt} 
        \centering
        \includegraphics[width=\columnwidth]{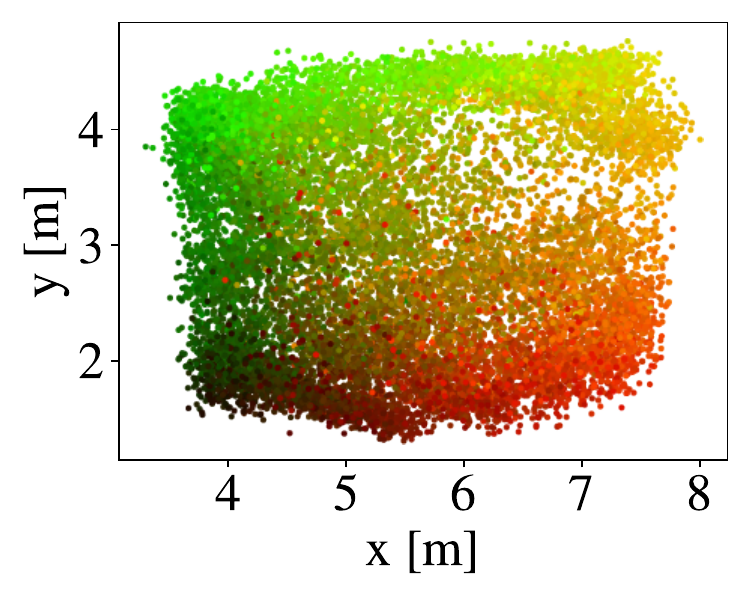}
    \end{minipage}\hfill
    \begin{minipage}[b]{.49\columnwidth}
        \hspace{-28pt} 
        \centering
        \includegraphics[width=\columnwidth]{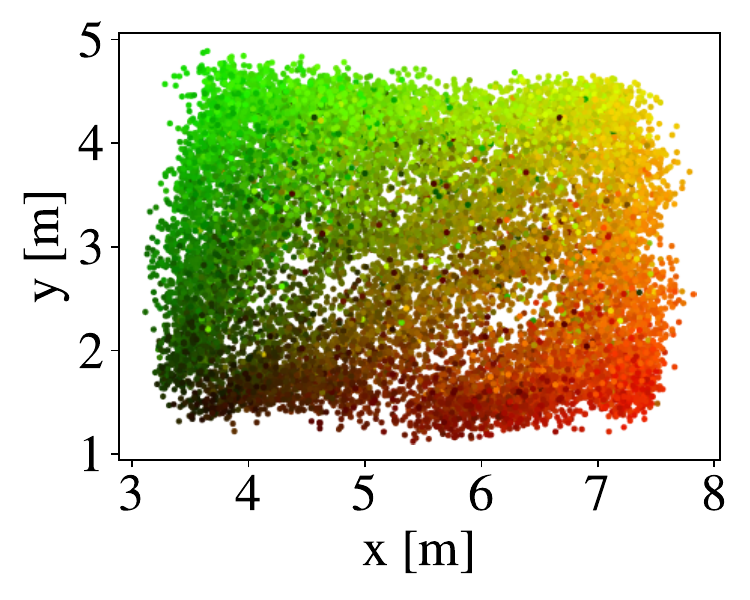}
    \end{minipage}
    \begin{minipage}[b]{.49\columnwidth}
        \hspace{-20pt} 
        \centering
        \includegraphics[width=\columnwidth]{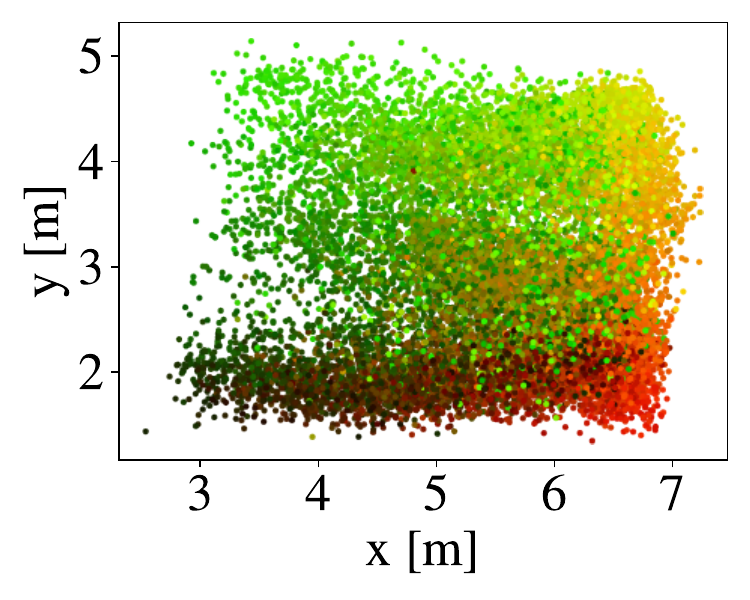}
    \end{minipage}
    \caption{ Visualization of the channel charts for configurations (left to right):  $st\_0$,  $st\_1$, $ac\_2\_s$, $ac\_2\_l$. Reference in Fig.~\ref{fig:cc_window_sizes}.}
    \label{fig:cc_ablation}
\end{figure*}

In this experiment, we vary the number of links between the nodes. We test the following configurations: 
single-transmitter setups $st\_\rho$, where $\rho$ is the transmitting node and the all connected mesh, where we utilize fewer nodes:
$ac\_5$ = [0,1,3,4,5], $ac\_4$ = [4,3,1,0], $ac\_3$ = [3,1,0], $ac\_2\_s$ = [3,0], and $ac\_2\_l$ = [2,5]. 
Fig.~\ref{fig:environment} shows the $st\_0$ setup in orange, using only links to node 0. 
We train \ac{pcc} with a 10\,s window. Fig.~\ref{fig:cc_ablation} and Tab.~\ref{tab:results_ablation} show the results, including the "full" mesh.

As before, supervised \ac{fp} outperforms \ac{pcc}. In the "full" mesh, \ac{fp} achieves CE90 = 0.31\,m/0.23\,m and MAE = 0.17\,m/0.13\,m (P1/P2), while \ac{pcc} reaches CE90 = 0.50\,m/0.42\,m and MAE = 0.28\,m/0.24\,m. Reducing the number of nodes decreases accuracy. With 5 nodes ($ac\_5$), \ac{fp} reaches CE90 = 0.41\,m/0.31\,m, \ac{pcc} reaches 0.55\,m/0.49\,m. With 2 nodes ($ac\_2\_s$), \ac{fp} drops to 0.69\,m/0.57\,m and \ac{pcc} to 0.85\,m/0.74\,m.

In $ac\_2\_l$, both \ac{fp} and \ac{pcc} perform worse compared to $ac\_2\_s$, despite the same number of nodes. We suspect overfitting, where the network learns patterns in the training data that do not reflect true position. With only 2 nodes, spatial information is limited, reducing generalization. The \ac{cc} shows deformed geometry, with many points pushed to the right (fourth graph), unlike $ac\_2\_s$ (third graph), which aligns with the reference. With 5 or 4 nodes ($ac\_5$, $ac\_4$), \ac{pcc} still performs well, staying below CE90 = 0.6\,m for both targets. With 3 or 2 nodes, accuracy drops significantly.

In single-transmitter setups $st\_\rho$, \ac{fp} accuracy worsens compared to the "full" mesh (CE90 = 0.40--0.71\,m vs. 0.31\,m). \ac{pcc} performs similarly to the "full" mesh, with only 0.05--0.10\,m difference in most cases. The drop in \ac{fp} may result from the noisy distance matrix estimated by PDR rather than the node setup. However, $st\_1$ shows unusually high error, potentially due to hardware impairment of the employed low cost sensors. Interestingly, $st\_\rho$-setups use only 6 links, yet match the performance of $ac\_5$ with 10 links. In some cases, e.g., $st\_0$, $st\_3$), they even outperform $ac\_4$ despite the same number of links. This suggests spatial node distribution matters more than total link count.

\vspace{+0.1cm}

Overall, a trade-off exists between hardware cost and system complexity. A single-transmitter setup reduces synchronization effort and avoids \ac{tdma} scheduling but increases node count and deployment cost. A full mesh lowers node count but reduces recording rate due to required scheduling.

\section{Conclusion}%
\label{sec:conclusion}%

To enable passive localization of moving people with an \ac{uwb} mesh, we propose \ac{pcc}, which builds \ac{fp} models using only a few reference positions. Our \ac{pdr}-based \ac{pcc} approach achieves high accuracy, with CE90 up to 0.50\,m on unknown targets, showing strong generalization.

We evaluated different \ac{uwb} configurations, varying the number of nodes and their connections. We found that spatial distribution is as important as the number of links, creating a trade-off between hardware cost and link budget.

\bibliography{library.bib}
\bibliographystyle{IEEEtran}

\end{document}